# Artificial Intelligence Applications in Lean Startup Methodology: A Bibliometric Analysis of Research Trends and Future Directions

Parisa Omidmand[1], Rasam Dorri[2], Alireza Mozaffari[3], Saeid Ataei[4]

[1]Department of Economics, Texas Tech University
[2]Department of Marketing, University of California, Riverside
[3]Department of Finance and Business Economics, Fordham University
[4]Department of Systems Engineering, Stevens Institute of Technology

**Abstract**

This study presents a comprehensive bibliometric analysis of the emerging intersection between artificial intelligence (AI) and lean startup methodology. Using the PRISMA 2020 framework, we systematically analyzed 12 peer-reviewed articles published between 2010 and June 2025, sourced from the Scopus database. The analysis employed VOS viewer software to conduct co-authorship, keyword co-occurrence, and citation network analyses. Results reveal three distinct research clusters: operational integration of AI within startup experimentation processes, AI-enhanced learning systems for entrepreneurial contexts, and strategic implications of AI for uncertainty management in startups. The findings indicate a developing research domain characterized by fragmented authorship networks, limited international collaboration, and geographic concentration in developed economies, particularly the United States and Germany. Key research themes include business model innovation, iterative methods, and machine learning applications, with artificial intelligence serving as a bridging concept across thematic clusters. The analysis identifies significant research gaps in ethical considerations, cross-cultural validation, and empirical testing of AI-enabled lean startup frameworks. While current research demonstrates growing interest in AI integration within entrepreneurial experimentation, the field requires enhanced theoretical consolidation, methodological rigor, and interdisciplinary collaboration to achieve practical relevance and academic maturity. This study contributes to the emerging discourse on digital entrepreneurship by providing a systematic overview of research trends and identifying priority areas for future investigation at the intersection of AI and lean startup methodologies.

# 1. Introduction

The convergence of artificial intelligence (AI) technologies with entrepreneurial methodologies represents one of the most significant developments in contemporary business innovation, fundamentally reshaping how startups approach market validation, product development, and strategic decision-making [24]. Lean startup methodology, pioneered by [14], has emerged as the dominant framework for entrepreneurial experimentation, emphasizing rapid prototyping, validated learning, and iterative development cycles to minimize resource waste and accelerate market entry. Simultaneously, advances in machine learning, predictive analytics, and automated decision-making systems have created unprecedented opportunities for enhancing entrepreneurial processes through data-driven insights and intelligent automation [3, 17].

The integration of AI within lean startup practices promises to address several fundamental challenges that characterize early-stage venture development. Traditional lean startup approaches rely heavily on human judgment and intuitive pattern recognition for interpreting market feedback and making pivoting decisions, processes that are inherently subject to cognitive biases and limited analytical capacity [5]. To overcome this, advanced AI techniques like sentiment analysis can automatically process vast amounts of social media data to gauge public opinion, providing startups with real-time market feedback that is less subject to human cognitive biases [17]. This capacity for objective data analysis extends to a deeper understanding of end-users. A compelling example is the 'Human Digital Twin', a dynamic, virtual model that captures an individual's unique preferences and behavioral patterns, thereby enabling the prediction of needs and the personalization of interactions [20].

The integration of AI technologies to lean startup methodology offers the potential to accelerate the build-measure-learn cycle through sophisticated data analysis, predictive modeling, and automated hypothesis testing, while improving the accuracy of market validation decisions. This integration, however, presents a double-edged sword, because "While Al can enhance innovation, it can also compromise autonomous reasoning if adopted uncritically" [21]. Consequently, while AI offers powerful analytical capabilities, its successful application within entrepreneurial contexts demands careful model selection, rigorous data handling, and a realistic grasp of complex market dynamics [18]. These requirements bring new complexities and theoretical challenges that call for systematic investigation. The deterministic nature of algorithmic decision-making may conflict with the experimental flexibility that characterizes successful entrepreneurial adaptation, while the data requirements for effective AI implementation may exceed the resource constraints typical of early-stage ventures [28]. Furthermore, the application of AI within entrepreneurial contexts raises important questions about the balance between automated efficiency and human creativity in innovation processes [31], as well as emergent power asymmetries that arise when AI systems retain comprehensive records of entrepreneurial decisions while human memory naturally fades [19].

Despite the growing practical interest in AI-enabled entrepreneurship, the academic literature examining this intersection remains fragmented and theoretically underdeveloped. Existing research spans multiple disciplines, including computer science, management, and entrepreneurship, but lacks systematic integration of perspectives and empirical validation of proposed frameworks [7]. The absence of comprehensive literature reviews examining this emerging domain limits both theoretical development and practical guidance for entrepreneurs seeking to leverage AI technologies within lean startup methodologies.

The rapid pace of AI advancement requires systematic tracking of research trends to guide future work. Bibliometric analysis is well-suited for mapping intellectual structure in emerging interdisciplinary domains. Unlike narrative reviews that depend on subjective synthesis, bibliometric methods use quantitative techniques to reveal structural patterns, identifying invisible colleges, research fronts, and knowledge flows between communities [4, 23]. For a field like AI-enabled lean startup methodology, where disciplinary boundaries remain unclear and terminology unstandardized, this approach reveals where intellectual silos have formed and which theories are being drawn upon, insights difficult to obtain through traditional reviews [23].

Understanding how AI integrates with lean startup methodology has implications beyond academia. It informs innovation policy, entrepreneurial education, and economic development, especially as accessible AI tools and lean practices shape competitive advantage at both venture and regional levels [2]. Finally, the ethical and societal dimensions of AI-driven entrepreneurial experimentation warrant close attention to ensure responsible innovation that benefits broader stakeholder communities [29].

This study addresses the identified research gap by conducting a comprehensive bibliometric analysis of the literature examining AI applications within lean startup methodology. Using the PRISMA 2020 framework for systematic review methodology, we analyzed peer-reviewed articles published between January 2010 and June 2025 to map the intellectual structure, identify key research themes, and examine collaborative patterns within this domain. The analysis employs advanced bibliometric techniques, including co-authorship network analysis, keyword co-occurrence mapping, and citation pattern examination, to provide objective insights into research trends and theoretical development.

The primary objectives of this investigation are threefold. First, we seek to map the current state of knowledge regarding AI integration within lean startup practices, identifying key authors, institutions, and publication outlets that are shaping discourse in this domain. Second, we aim to identify the primary research themes and theoretical clusters that characterize existing scholarship, examining how different disciplinary perspectives are contributing to the understanding of AI-enabled entrepreneurship. Third, we endeavor to identify research gaps and future directions that could guide subsequent empirical investigation and theoretical development in this field [16].

The findings of this bibliometric analysis contribute to the emerging literature on digital entrepreneurship by providing the first systematic overview of research at the intersection of AI and lean startup methodology. For academic researchers, this study offers a roadmap for future investigation by identifying underexplored themes and methodological gaps that require attention. For practitioners and policymakers, the analysis provides insights into the current state of knowledge and evidence-based guidance for supporting AI-enabled entrepreneurial innovation.

The remainder of this paper is structured as follows. Section 2 details the methodology employed for the bibliometric analysis, including search strategy, inclusion criteria, and analytical techniques. Section 3 presents the results of the bibliometric investigation, examining authorship patterns, thematic clustering, and research evolution. Section 4 discusses the theoretical and practical implications of the findings, identifies research gaps, and proposes directions for future investigation. Section 5 concludes with a synthesis of key insights and recommendations for advancing research in this emerging domain.

## 2. Methodology

This study employed a systematic bibliometric analysis to examine the application of artificial intelligence in lean startup methodology. The research design followed the Preferred Reporting Items for Systematic Reviews and Meta-Analyses (PRISMA) 2020 guidelines, which provide a comprehensive framework for conducting transparent and reproducible systematic reviews [13]. The PRISMA framework ensures methodological rigor by establishing standardized procedures for literature identification, screening, and selection, thereby minimizing bias and enhancing the reliability of systematic reviews [10]. Given the interdisciplinary nature of AI applications in entrepreneurial contexts, PRISMA's structured approach was particularly valuable for maintaining consistency throughout the review process and enabling replication of the methodology.

The Scopus database was selected as the primary data source for this bibliometric analysis due to its superior coverage of multidisciplinary scholarly literature and robust bibliometric capabilities. Scopus indexes over 42,000 active journals and provides comprehensive citation tracking, making it particularly suitable for bibliometric studies spanning multiple disciplines [11, 1]. Compared to Web of Science, Scopus demonstrates broader coverage of computer science and business literature, which are central to this investigation [6]. While Google Scholar offers extensive coverage, its inclusion of grey literature and variable quality control makes it less appropriate for systematic bibliometric analysis requiring consistent academic standards [8].

The systematic search was conducted using a comprehensive Boolean query applied to title, abstract, and keywords fields: ("artificial intelligence" OR "machine learning" OR "deep learning" OR "predictive analytics" OR "natural language processing" OR "computer vision" OR "automated decision making" OR "intelligent systems" OR "AI" OR "ML") AND ("lean startup" OR "minimum viable product" OR "MVP" OR "pivot" OR "validated learning" OR "customer development" OR "build-measure-learn" OR "startup methodology" OR "entrepreneurial experimentation" OR "business model validation"). This search strategy captured the essential terminology from both AI and lean startup domains while accommodating variations in academic nomenclature. The initial search yielded 1,437 records, which were subsequently refined through systematic filtering criteria aligned with PRISMA guidelines. Language restrictions limited results to English publications (n = 1,384), followed by subject area filtering to include Business, Management and Accounting, Computer Science, Engineering, Economics, Econometrics, and Finance (n = 983). Document type restrictions to articles and reviews reduced the corpus to 343 publications, with temporal boundaries set from 2010 to 2025, yielding 296 eligible records.

The screening process followed PRISMA's two-stage approach, beginning with title and abstract evaluation. Of the 296 records, 274 were excluded as irrelevant, domain-specific, or lacking substantive integration of AI and lean startup concepts. The remaining 22 reports underwent full-text assessment, resulting in the exclusion of 10 studies that discussed AI or lean startup methodology in isolation without demonstrating methodological integration. This rigorous screening process culminated in 12 studies meeting the inclusion criteria for final analysis. While 12 studies may seem modest, this figure reflects the state of a field at the intersection of two evolving domains. As Donthu et al. [4] observe, bibliometric analyses of emerging fields serve a distinct purpose: establishing baseline intellectual structures before a field reaches critical mass. This analysis is therefore a foundational mapping exercise that documents current knowledge and identifies priorities for future development.

The Bibliometric analysis and network visualizations were conducted using VOSviewer software version 1.6.20 [15]. VOSviewer was employed to generate co-authorship networks, keyword co-occurrence maps, bibliographic coupling analyses, and citation network visualizations, enabling systematic examination of collaborative patterns, thematic clustering, and intellectual structure within the research domain. The complete selection process is illustrated in Figure 1: PRISMA Flow Diagram, which provides a transparent account of the systematic filtering and decision-making procedures employed throughout this bibliometric investigation.

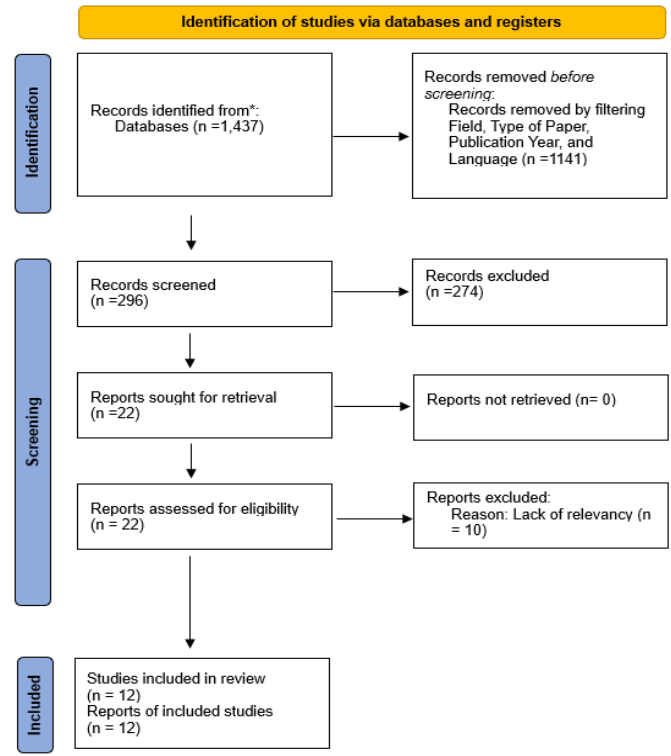

**Figure 1.** PRISMA 2020 flow diagram illustrating the systematic process of study identification, screening, and selection

## 3. Results

*3.1 Bibliometric Analysis Overview*

The systematic review identified 12 studies that demonstrate the emerging integration of artificial intelligence technologies within lean startup methodologies. The analysis reveals a nascent but growing research field, with publications spanning from 2017 to 2025, indicating the relatively recent convergence of these two domains. The temporal distribution suggests an acceleration in research interest, with an increase in publications between 2021 and 2023, as demonstrated in Figure 2, potentially reflecting the growing practical application of AI technologies in entrepreneurial contexts.

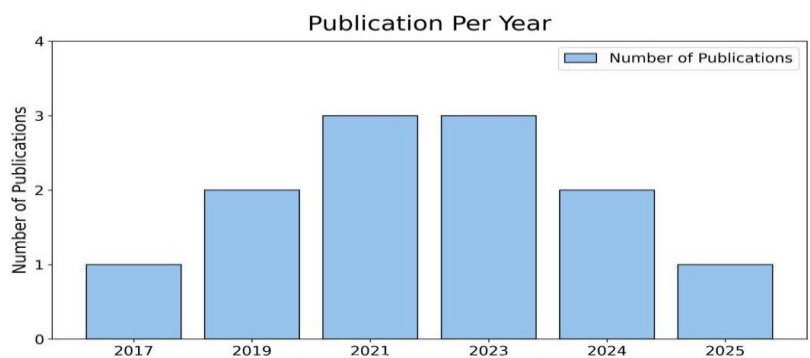

**Figure 2.** Temporal distribution of publications

*3.2 Author and Publication Patterns*

The intellectual structure of the field was mapped using a bibliographic coupling analysis of authors, as visualized in Figure 3. This technique connects authors who cite a common set of references, revealing shared intellectual foundations. The analysis was filtered to include only authors with a minimum of 10 citations to focus on the most impactful contributors.

The network is characterized by a linear structure with several distinct clusters (indicated by color), suggesting the field is organized into a few related, but not highly overlapping, research streams. For example, Dellermann, D., and Parida, V., form one conceptual cluster (green-yellow), indicating their work is built upon a similar body of literature. A separate cluster (red-purple) includes Alasim, F., and Brecht, P., suggesting they are engaged in a different research conversation. The author Fouss, F., is positioned centrally, signifying that his work may bridge these intellectual streams by drawing upon references from multiple clusters. This structure implies that the field is developing along distinct conceptual paths with low direct author collaboration.

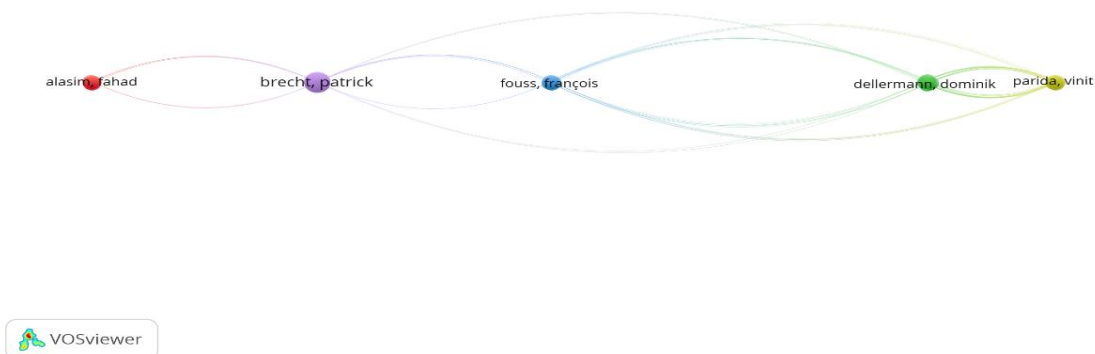

**Figure 3.** Bibliographic coupling of authors with a minimum of 10 citations

The performance metrics for the authors included in the network analysis are detailed in Table 1. The table's columns allow for a simultaneous analysis of author productivity (the 'Documents' column) and influence (the 'Citations' column), and Total Link Strength (TLS).

An examination of the 'Documents' column reveals that productivity in the field is low. Among the 18 most-cited authors, only Brecht Patrick and Hahn Carsten H. have more than one publication. Their total of 13 citations is distributed between their two papers (one with 8 citations and the other with 5). This finding reinforces the conclusion that the field lacks a core group of researchers consistently publishing multiple works on this topic.

The 'Citations' column highlights the nature of scholarly influence in this domain. The data shows that influence is not driven by individual scholars but is concentrated in a few highly-cited, multi-authored papers. For example, the four authors with the highest citation count (74) are co-authors on a single publication [29]. This pattern of shared, identical citation counts repeats for the other top author groups. This emphasizes that while the number of collaborations is low overall, the most impactful work in this developing field is the result of collaborative efforts.

This pattern also indicates that breakthrough contributions emerge from intensive small-team collaboration rather than broad network cooperation. The modest impact of even the most influential work, when compared to established fields, points to the early stage of knowledge accumulation in this interdisciplinary area. Finally, the 'Total Link Strength' column reveals that the most-cited authors form the field's intellectual core, while the most productive authors are more conceptually peripheral.

**Table 1.** Documents, Citations, and Link Strength

| Author | Documents | Citations | Total Link Strength |
|---|---|---|---|
| **Dellermann, Dominik** | 1 | 74 | 359 |
| **Ebel, Philipp** | 1 | 74 | 359 |
| **Leimeister, Jan Marco** | 1 | 74 | 359 |
| **Lipusch, Nikolaus** | 1 | 74 | 359 |
| **Jovanovic, Marin** | 1 | 11 | 255 |
| **Parida, Vinit** | 1 | 11 | 255 |
| **Sjodin, David** | 1 | 11 | 255 |
| **Thomson, Linus** | 1 | 11 | 255 |
| **Fouss, François** | 1 | 17 | 240 |
| **Hermans, Julie** | 1 | 17 | 240 |
| **Lecron, Fabian** | 1 | 17 | 240 |
| **Raneri, Santo** | 1 | 17 | 240 |
| **Alasim, Fahad** | 1 | 20 | 160 |
| **Nagadi, Khalid** | 1 | 20 | 160 |
| **Rabelo, Luis** | 1 | 20 | 160 |
| **Shiradkar, Sayli** | 1 | 20 | 160 |
| **Brecht, Patrick** | 2 | 13 | 93 |
| **Hahn, Carsten H.** | 2 | 13 | 93 |

*3.3 Keyword Co-occurrence and Thematic Clustering*

To map the primary research themes, a keyword co-occurrence analysis was conducted on all keywords appearing at least twice. The frequency and network metrics for these keywords are detailed in Table 2. The frequency data reveals the prominence of the field's core terms, "artificial intelligence" repeated 4 times and "lean startup" repeated 3 times, confirming their centrality. However, the relatively low occurrence frequencies across all keywords (maximum 4) indicate conceptual dispersion rather than thematic convergence, suggesting that researchers have not yet established standardized terminology or unified theoretical frameworks.

**Table 2.** Keyword Co-occurrence Metrics

| Keyword | Occurrences | Total Link Strength |
|---|---|---|
| **Artificial Intelligence** | 4 | 10 |
| **Lean Startup** | 3 | 8 |
| **Business Model** | 3 | 7 |
| **Business Experiments** | 2 | 6 |
| **Iterative Methods** | 2 | 6 |
| **Learning Systems** | 3 | 6 |
| **Business Models** | 2 | 5 |
| **Uncertainty** | 2 | 4 |
| **E-learning** | 2 | 3 |
| **Machine Learning** | 2 | 3 |
| **Entrepreneurship** | 3 | 2 |

To visualize the relationships between these concepts, a keyword co-occurrence network was generated, as shown in Figure 4. The network structure reveals three distinct clusters (indicated by color) with varying densities and connectivity patterns. The spatial separation between clusters indicates conceptual boundaries, while the cluster sizes suggest differential research attention across thematic areas. Within this structure, "artificial intelligence" (Occurrences = 4, Total Link Strength = 10) serves as a critical bridging node, connecting the different research streams.

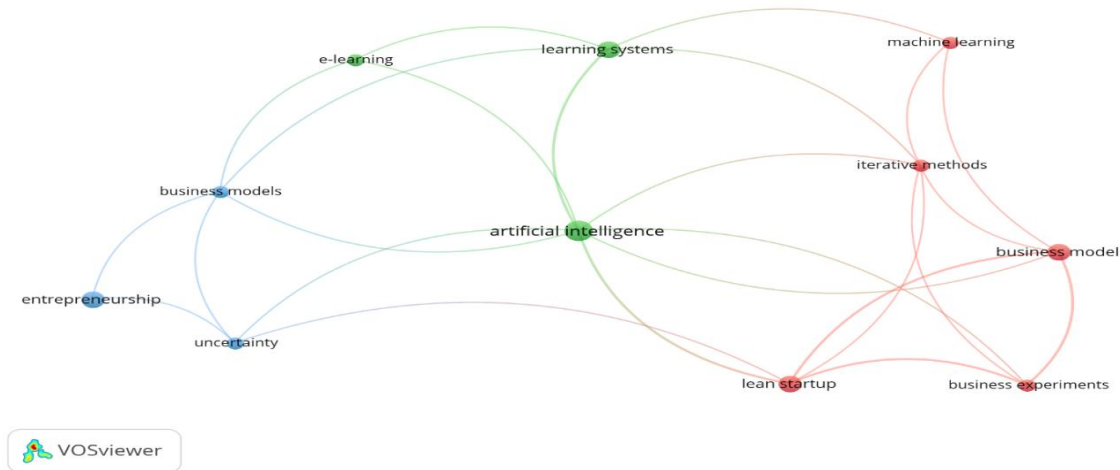

**Figure 4.** Keyword co-occurrence network

Operational Integration of AI Cluster (Red):

This cluster encompasses core lean startup concepts, including "lean startup," "business model," "business experiments," "iterative methods," and "machine learning," indicating a strong focus on the operational integration of AI within startup experimentation processes. This cluster demonstrates the highest internal connectivity in the network, confirmed by the high Total Link Strength of its central node, "lean startup" (TLS = 8)."

AI-Enhanced Learning Systems Cluster (Green):

This cluster centers on learning and educational applications, featuring "artificial intelligence," "learning systems," and "e-learning," suggesting a significant research stream focused on AI's role in enhancing learning processes within entrepreneurial contexts. The positioning of artificial intelligence as a bridge between clusters indicates its cross-cutting relevance, supported by its network-leading Total Link Strength (TLS = 10). This cluster's moderate density suggests emerging but not fully developed connections between AI technologies and learning processes in entrepreneurial contexts.

Strategic and Theoretical Dimensions of AI Cluster (Blue):

This Cluster emphasizes strategic and theoretical dimensions, incorporating "entrepreneurship," "business models," and "uncertainty," reflecting research attention toward the broader implications of AI integration for entrepreneurial strategy and risk management. A key analytical insight is the peripheral position of "uncertainty," which is confirmed by its low link strength (TLS = 4). This relative isolation suggests that risk-related research remains underdeveloped, representing a significant research gap, as uncertainty management constitutes a core challenge in entrepreneurial experimentation.

*3.4 Journal Distribution and Intellectual Foundations*

The analysis of publication sources reveals a significant diversity in journal outlets, spanning business, technology, and interdisciplinary domains. As detailed in Table 3, the 12 articles in this review were published across 12 different journals. The absence of a dominant journal, where no single outlet has published more than one article, signifies that the field has not yet established preferred publication venues or clear disciplinary boundaries, a common characteristic of a developing research area.

**Table 3.** Publication Outlets and Performance Metrics

| Source | Documents | Citations | Total Link Strength |
|---|---|---|---|
| **Strategy Science** | 1 | 5 | 15 |
| **Electronic Markets** | 1 | 5 | 13 |
| **International Review of Economics and Finance** | 1 | 5 | 12 |
| **Technovation** | 1 | 11 | 11 |
| **International Journal of Entrepreneurial Behavior & Research** | 1 | 17 | 10 |
| **Artificial Intelligence for Engineering Design, Analysis and Manufacturing** | 1 | 8 | 8 |
| **Technology Innovation Management Review** | 1 | 5 | 7 |
| **IEEE Transactions on Engineering Management** | 1 | 1 | 4 |
| **Information and Software Technology** | 1 | 6 | 4 |
| **Information (Switzerland)** | 1 | 20 | 2 |
| **Case Journal** | 1 | 0 | 0 |
| **IEEE Software** | 1 | 7 | 0 |

To understand the relationships between these publication outlets, a bibliographic coupling analysis was performed, as shown in Figure 5. This map visualizes which journals in the sample cite similar bodies of literature. The network demonstrates moderate clustering, with academic management journals (e.g., Strategy Science, International Review of Economics and Finance) forming a distinct cluster separate from technology-focused publications (e.g., Technovation, Information and Software Technology). This bifurcation pattern suggests that research in this domain may be developing along parallel disciplinary lines, potentially limiting the cross-pollination of ideas between technical and managerial perspectives.

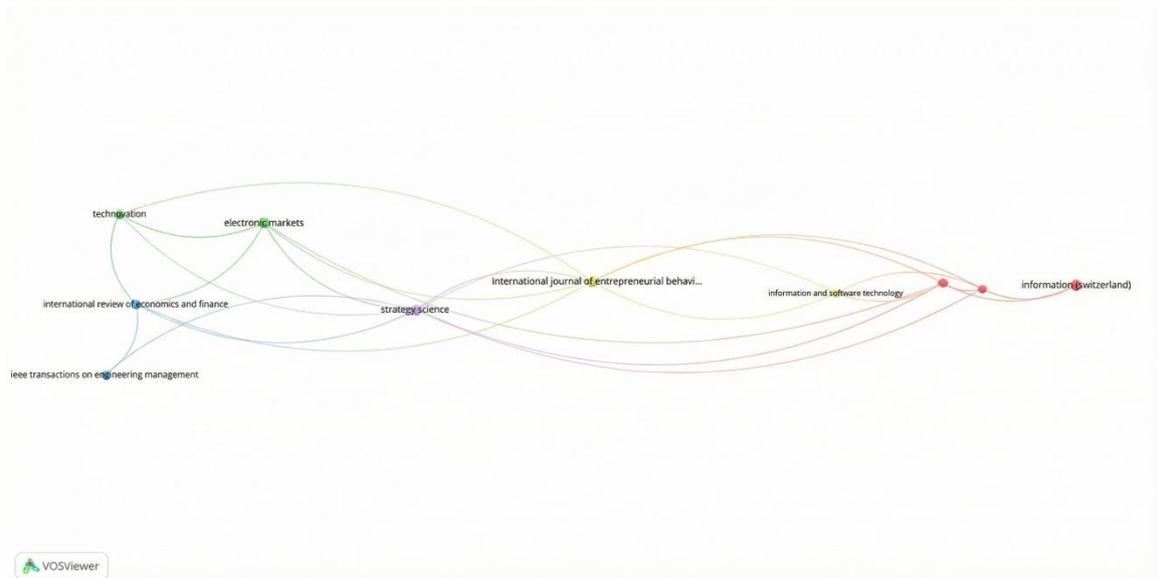

**Figure 5.** Network visualization of journal publication outlets

To identify the field's intellectual foundations, a co-citation analysis was conducted, with the resulting map visualized in Figure 6. This map shows the established journals that are most frequently cited together by the 12 articles in the sample. The analysis clearly reveals that the research is not built on a single tradition but is instead shaped by three distinct intellectual domains:

The Entrepreneurship Cluster (Blue): The largest cluster is centered on core entrepreneurship and innovation journals, such as Small Business Economics and Entrepreneurship Theory and Practice. This indicates that the research is firmly rooted in established entrepreneurship theory.

The General Management Cluster (Red): A second cluster consists of top-tier general management and strategy journals, including Academy of Management Review and Long-Range Planning. This shows that scholars are drawing on foundational theories from the broader management discipline to frame their work.

The Information Systems Cluster (Green): The third cluster is composed of leading Information Systems (IS) journals like MIS Quarterly and Decision Support Systems. This highlights the critical role of technological and systems-based theories in this research area.

Taken together, the co-citation map demonstrates that the intersection of AI and lean startup is a truly interdisciplinary domain, synthesizing knowledge from three distinct fields: entrepreneurship, general management, and information systems.

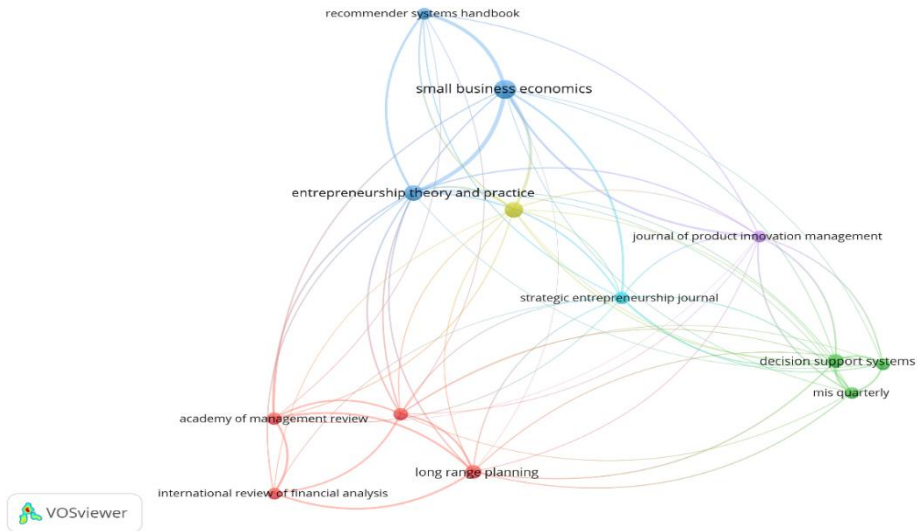

**Figure 6.** Source co-citation network

*3.5 Geographic Distribution and Institutional Collaboration*

To examine the geographic distribution of research contributions, a co-authorship analysis by country was conducted using VOSviewer software. The analysis extracted country information from author institutional affiliations in the twelve publications, counting the number of documents with at least one author from each country.

The geographic analysis of author affiliations reveals a research landscape dominated by developed economies, as detailed in Table 4 and visualized in the choropleth map in Figure 7. The United States is the leading contributor, accounting for 4 of the 23 total authorial contributions (17.4%), followed by Germany with 3 contributions (13%). This concentration aligns with the leadership of these regions in both AI research and mature entrepreneurial ecosystems. Notably, there is a strong showing from Nordic countries (Sweden, Norway, Denmark, and Finland), which collectively account for 6 contributions, suggesting a regional strength in this area. This pattern, with a notable absence of contributions from many developing economies, indicates that AI-lean startup research requires advanced technological infrastructure and mature entrepreneurial ecosystems that facilitate the convergence of artificial intelligence capabilities with startup methodologies.

**Table 4.** Geographic Distribution of Research Contributions

| Country | Number of Documents |
| --- | --- |
| **United States** | 4 |
| **Germany** | 3 |
| **Norway** | 2 |
| **Sweden** | 2 |
| **India** | 2 |
| **Canada** | 2 |
| **Denmark** | 1 |
| **Finland** | 1 |
| **Italy** | 1 |
| **Saudi Arabia** | 1 |
| **United Kingdom** | 1 |
| **Switzerland** | 1 |
| **Belgium** | 1 |
| **China** | 1 |

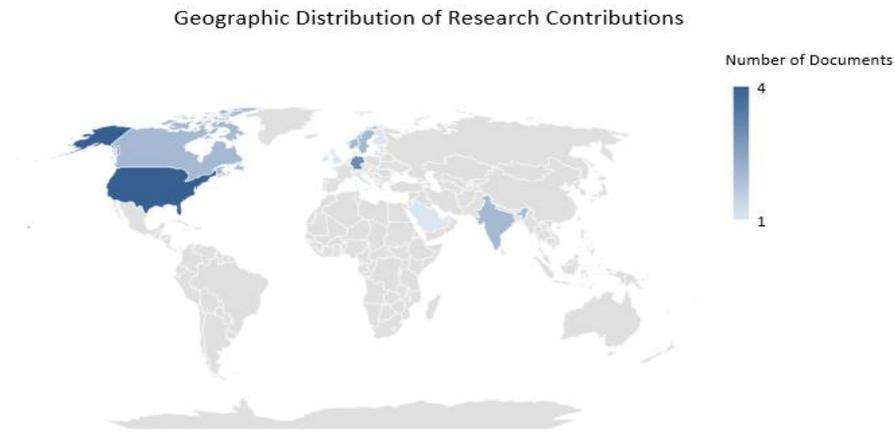

**Figure 7.** Geographic map of global research contributions

At the institutional level, collaboration patterns are limited. The institutional affiliation network, Figure 8, illustrates a fragmented research landscape with minimal cross-cluster and international cooperation. Most research appears to be conducted within distinct national or regional boundaries, indicating missed opportunities for the cross-cultural validation of AI-lean startup frameworks. This fragmentation suggests that current research findings may reflect regional biases rather than universally applicable principles.

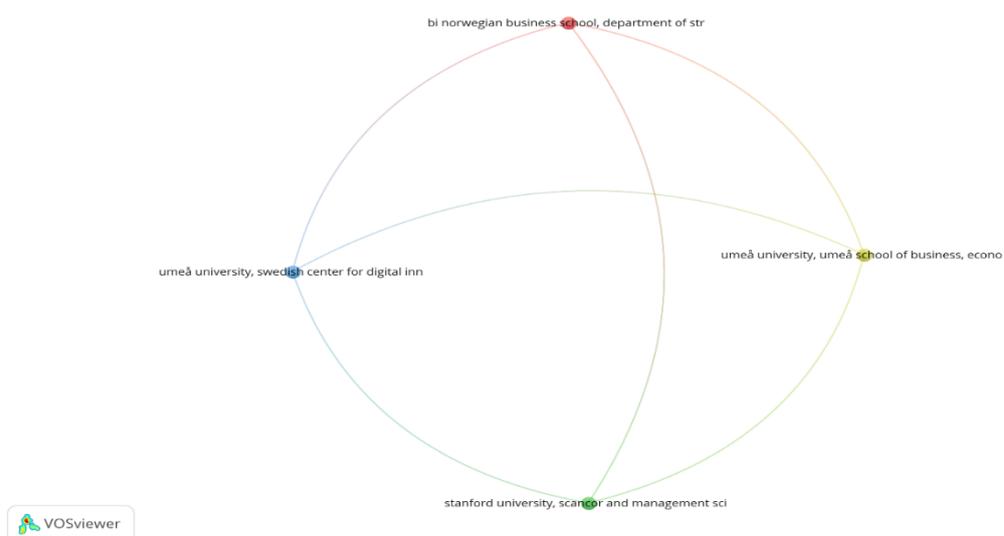

**Figure 8.** Institutional collaboration network based on author affiliations

*3.6 Research Maturity and Development Patterns*

The bibliometric indicators collectively reveal that AI applications in lean startup methodology represent an emerging research domain in a pre-paradigmatic stage, characterized by early conceptual development rather than empirical consolidation. The fragmented authorship networks documented in Section 3.2, combined with low overall citation counts and the conceptual dispersion of keywords identified in Section 3.3, all point to a field that has not yet established theoretical convergence or dominant paradigms.

The field's formative theoretical foundations are evident in the co-citation network of cited authors, visualized in Figure 9. This map reveals the distinct intellectual traditions upon which this emerging research is built. A large, dense cluster (red) represents the foundational entrepreneurship literature, including key scholars of the Lean Startup methodology (Ries, E.; Blank, S.), Effectuation theory (Sarasvathy, S.D.), and the Business Model Canvas (Osterwalder, A.; Pigneur, Y.). This core is loosely connected to a second group of scholars (green) focused on Strategy and Business Models (e.g., Eisenhardt, K.M.; Amit, R.; Zott, C.) and a third, more recent stream centered on Modern Digital Strategy (Parida, V., Sjödin, D.). The relative separation and weak linkages between these intellectual clusters provide strong visual evidence that while researchers are drawing upon established theories, they have not yet integrated them into a single, unified framework for the specific context of AI in lean startups.

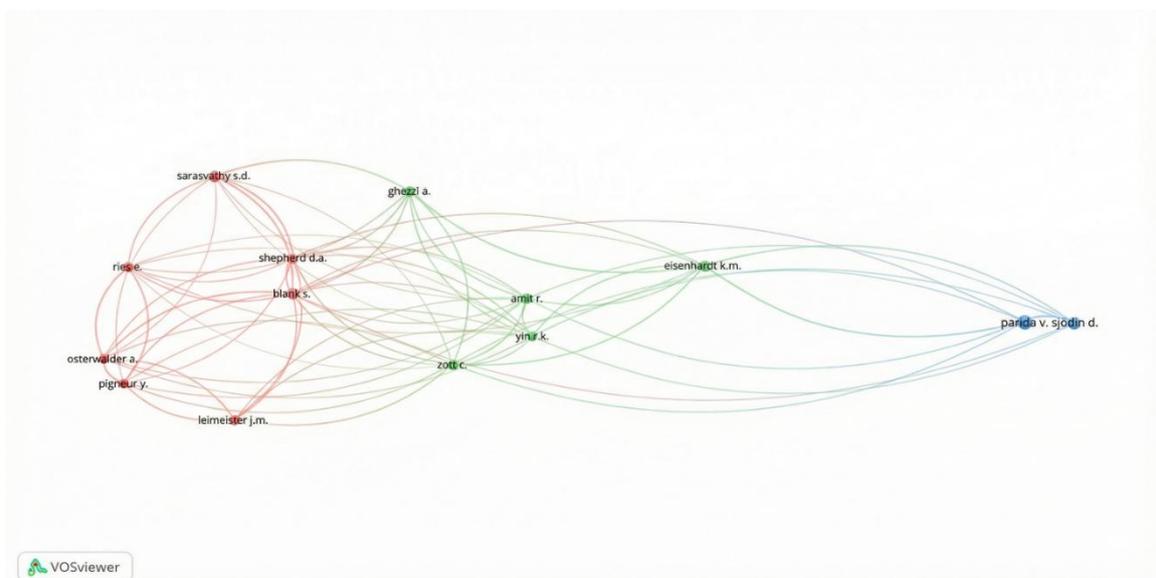

**Figure 9.** Co-citation network of cited authors.

Despite its nascent stage, the field shows a clear evolutionary trajectory, as illustrated in the temporal overlay of the keyword network in Figure 10. The analysis demonstrates a progression from earlier, foundational concepts (e.g., "learning systems" and "iterative methods" in blue, c. 2018-2020) toward more recent, strategic themes like "business model design" and "uncertainty" (yellow, c. 2022-2024). However, this evolution also reveals a significant research gap: the limited occurrence of advanced AI concepts such as "deep learning" or "predictive analytics" indicates that research has not yet fully exploited the potential of sophisticated AI technologies within lean startup contexts, representing a significant opportunity for future investigation.

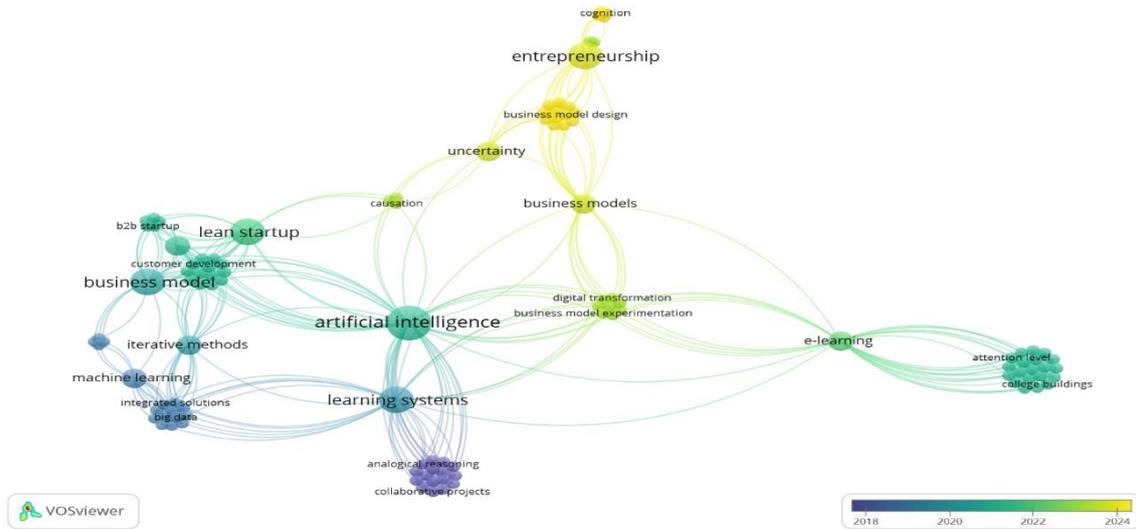

**Figure 10.** Temporal overlay of the keyword co-occurrence network

## 4. Discussion

*4.1 Theoretical Contributions and Implications*

The bibliometric analysis reveals that the integration of artificial intelligence within lean startup methodology represents an emergent theoretical domain characterized by conceptual fragmentation rather than paradigmatic consolidation. The identification of three distinct research clusters in Figure 4, including operational integration, learning systems, and strategic implications, suggests that scholars are approaching AI-enabled lean startup practices from multiple theoretical perspectives without establishing unified conceptual frameworks. This fragmentation reflects the inherent complexity of bridging computational intelligence with entrepreneurial experimentation, where traditional lean startup principles must be reconceptualized to accommodate AI's predictive capabilities and automated decision-making processes.

The prominence of "business model" and "iterative methods" as central nodes in the keyword network indicates that researchers are primarily focusing on how AI can enhance existing lean startup processes rather than fundamentally transforming entrepreneurial methodologies. This incremental approach suggests that the field is in its initial adaptation phase, where AI is viewed as a tool for optimization rather than a catalyst for methodological revolution. However, the isolated positioning of "uncertainty" in the keyword analysis reveals a critical theoretical gap, as uncertainty management represents a core challenge in both AI implementation and lean startup validation processes.

The temporal analysis shows an evolution from basic AI applications to more advanced integration frameworks, indicating a progression from proof-of-concept studies to systematic methodology development. This trend aligns with broader patterns in digital entrepreneurship research, where technological capabilities increasingly influence fundamental business practices [12]. However, the limited occurrence of advanced AI concepts such as machine learning algorithms or predictive analytics suggests that research has yet to fully explore the transformative potential of sophisticated AI technologies in entrepreneurial settings [22]. In this context, the examples discussed in the Introduction, such as sentiment-analysis-driven market feedback and Human Digital Twin systems, directly correspond to the "operational integration" and "AI-enhanced learning systems" clusters identified in the bibliometric analysis. Their limited visibility in keyword frequency and network density further supports the conclusion that such advanced applications remain conceptually influential but are still underdeveloped empirically within the current literature.

*4.2 Methodological Landscape and Research Gaps*

The methodological analysis reveals significant limitations in current research approaches, with most studies adopting conceptual or case-based methodologies rather than empirical validation of AI-enabled lean startup frameworks. The fragmented authorship networks and limited collaborative patterns suggest that research is conducted in disciplinary silos, potentially constraining the development of integrated theoretical perspectives that bridge computer science, entrepreneurship, and management science.

The geographic concentration of research in developed economies with mature startup ecosystems (the United States, Germany, and Nordic countries) raises questions about the generalizability of findings to diverse entrepreneurial contexts. This limitation is particularly significant given that AI adoption patterns and entrepreneurial practices vary substantially across different economic and cultural environments [2]. The absence of research from emerging economies, where resource constraints might necessitate different AI integration strategies, represents a critical gap in understanding the universal applicability of AI-enabled lean startup methodologies.

Furthermore, the analysis reveals insufficient attention to ethical and regulatory dimensions of AI integration within startup contexts. The keyword analysis shows no occurrence of terms related to AI ethics, data privacy, or algorithmic bias, despite these concerns being central to responsible AI implementation [9]. This omission suggests that current research prioritizes technical and operational aspects while neglecting the broader societal implications of AI-enabled entrepreneurial experimentation.

*4.3 Practical Implications for Entrepreneurs and Practitioners*

The research synthesis offers several practical insights for entrepreneurs seeking to integrate AI technologies within lean startup methodologies. The clustering analysis of Keyword co-occurrence network suggests that successful AI integration requires simultaneous attention to operational efficiency, organizational learning capabilities, and strategic positioning. This multi-dimensional approach indicates that entrepreneurs cannot simply overlay AI tools onto existing lean processes but must develop integrated capabilities that span technical implementation, learning system design, and strategic decision-making.

The prominence of "business experiments" and "iterative methods" in the keyword network suggests that AI's primary value lies in enhancing experimental rigor and accelerating iteration cycles rather than replacing human judgment in entrepreneurial decision-making. This finding aligns with recent practitioner literature emphasizing AI's role in augmenting rather than automating entrepreneurial cognition [30]. Practitioners should therefore focus on developing AI capabilities that enhance data collection, pattern recognition, and hypothesis testing while preserving the creative and intuitive aspects of entrepreneurial experimentation [31].

The limited research on uncertainty management represents a significant challenge for practitioners, as AI implementation inherently introduces new forms of uncertainty related to algorithmic reliability, data quality, and model interpretability. Entrepreneurs must develop capabilities to manage both traditional market uncertainties and AI-specific technical uncertainties, requiring new frameworks for risk assessment and mitigation strategies [24].

*4.4 Implications for Academic Research and Future Directions*

The bibliometric analysis identifies several priority areas for future research development. First, the field requires more rigorous empirical studies that validate proposed AI-enabled lean startup frameworks through longitudinal analysis of startup performance outcomes. The current predominance of conceptual studies limits practical applicability and theoretical advancement, necessitating systematic empirical investigation of causal relationships between AI integration and entrepreneurial success metrics.

Second, future research should develop more sophisticated theoretical frameworks that address the fundamental tensions between AI's deterministic capabilities and lean startup's experimental philosophy [26]. This theoretical development requires interdisciplinary collaboration between computer scientists, entrepreneurship scholars, and management researchers to create integrated perspectives that transcend disciplinary boundaries [25].

Third, the geographic concentration of current research highlights the need for cross-cultural studies that examine how different institutional contexts influence AI adoption patterns within entrepreneurial ecosystems. Such research would enhance the external validity of findings and provide insights into context-specific implementation strategies for diverse entrepreneurial environments. The isolated positioning of uncertainty-related concepts in the keyword analysis suggests need for research addressing risk management frameworks for AI-enabled entrepreneurship. Future studies should investigate how entrepreneurs can balance AI's predictive capabilities with the inherent unpredictability of innovative ventures, developing new approaches to uncertainty assessment and mitigation [24].

*4.5 Limitations and Methodological Considerations*

This bibliometric analysis is subject to several limitations that constrain the generalizability of findings. The relatively small corpus of 12 studies reflects the early stage of the research domain but limits the statistical power of bibliometric techniques and may not capture the full diversity of research approaches within this emerging field. The exclusive focus on Scopus-indexed publications may have excluded relevant gray literature, conference proceedings, and practitioner-oriented publications that could provide additional insights into AI-enabled lean startup practices.

The temporal scope of the analysis (2010-2025) encompasses the emergence of both modern AI technologies and lean startup methodologies, but the concentration of publications in recent years limits the ability to identify long-term trends or theoretical development patterns. Future bibliometric analyses would benefit from longitudinal tracking as the field matures and accumulates a larger corpus of empirical studies.

## 5. Conclusion

The bibliometric analysis reveals that AI integration within lean startup methodology represents a promising but underdeveloped research domain characterized by conceptual fragmentation, methodological limitations, and significant opportunities for theoretical advancement. While current research demonstrates growing interest in AI-enabled entrepreneurial experimentation, the field requires more rigorous empirical validation, theoretical integration, and attention to ethical and contextual considerations to achieve practical relevance and academic maturity.

The identification of three distinct research clusters provides a foundation for future theoretical development, but the limited interconnections between clusters suggest the need for more integrative approaches that bridge operational, learning, and strategic perspectives. As AI technologies continue to evolve and entrepreneurial practices adapt to digital transformation, this research domain has the potential to generate significant theoretical insights and practical innovations that enhance entrepreneurial effectiveness and societal value creation.